\documentclass[utf8]{FrontiersinHarvard} 

\usepackage{url,hyperref,lineno,microtype,subcaption}
\usepackage[onehalfspacing]{setspace}

\def\keyFont{\fontsize{8}{11}\helveticabold }
\def\firstAuthorLast{Finke} 
\def\Authors{Justin D.\ Finke$^{1,*}$}

\def\Address{$^{1}$U.S. Naval Research Laboratory, Code 7653, Washington, DC 20375-5352 USA}

\def\corrAuthor{Justin D.\ Finke \\
U.S. Naval Research Laboratory \\
Code 7653 \\
4555 Overlook Ave. SW \\
Washington, DC 20375-5352, USA
}

\def\corrEmail{justin.d.finke.civ@us.navy.mil}

\newcommand{\g}{\gamma}

\newcommand{\psim}{\lower.5ex\hbox{$\; \buildrel \propto \over\sim \;$}}
\newcommand{\lbar}{\lower.0ex\hbox{$\; \buildrel
{\lower0.0ex \hbox{-}} \over\lambda  \;$}}

\newcommand{\tilx}{\tilde{x}}

\newcommand{\cm}{\mathrm{cm}}

\newcommand{\erg}{\mathrm{erg}}

\newcommand{\s}{\mathrm{s}}

\begin{document}
\onecolumn
\firstpage{1}

\title[Light Travel Time Effects in Blazar Flares]{Light Travel Time Effects in Blazar Flares} 

\author[\firstAuthorLast ]{\Authors} 
\address{} 
\correspondance{} 

\extraAuth{}

\maketitle

\begin{abstract}

I present a model for light travel time effects for emission from a plasma blob in a blazar jet.  This calculation could be incorporated into more complex models with particle acceleration and radiation mechanisms, but as presented here it is a agnostic as to these mechanisms.  This model includes light travel time effects for an expanding or contracting blob.  As an example, this model is applied to a flare observed by VERITAS and MAGIC from Mrk 421 in 2013; and to a flare observed by the {\em Fermi} Large Area Telescope from 3C 454.3 in 2010.  



\tiny
 \keyFont{ \section{Keywords:} BL Lacertae object: general, BL Lacertae objects:  individual (Mrk 421), quasars: general, quasars: individual (3C 454.3), gamma-rays: general, gamma-rays: galaxies, time-domain astronomy}
\end{abstract}

\section{Introduction}
\label{introsection}


Blazars are active galactic nuclei (AGN) with relativistic jets moving close to our line of sight.  They are routinely detected at all wavelengths, from radio to $\gamma$-rays, and often extremely variable at all wavelengths as well.  Consequently, a complete understanding of blazars requires a deep understanding of the time domain.  The {\em Fermi} Large Area Telescope (LAT) is extremely useful for this, since it monitors the entire sky in $\approx 30$\ MeV to 100 GeV $\gamma$-rays every 3 hours.  Observations by the LAT can be supplemented with observations by many other observatories, for instance by {\em Swift} in space in the optical and X-rays; and on the ground by MAGIC, H.E.S.S., and VERITAS at very-high energy (VHE) $\gamma$-rays, Very Large Baseline Array, TANAMI, Effelsberg, Owens Valley Radio Observatory in the radio, and SMARTS and the GASP-WEBT consortium in the optical (among others).  The loss of {\em Fermi} will be a major blow to the study of blazars, especially considering there is no planned observatory that can replace its capabilities.

Blazar variability is often analyzed in terms of flares, explained by particle acceleration and radiative cooling in a homogeneous one-zone nonthermal plasma ``blob''.  Extensive time-dependent modeling analysis on flares has been performed \citep[e.g.,][]{boettcher97,kirk98,chiaberge99,dermer02,joshi11,zacharias13,dotson15,zacharias23}.  Particle acceleration could be from diffusive shock or magnetic reconnection.  Leptonic radiation processes include synchrotron, typically at low energies, and Compton scattering of a number of radiation fields at high energies.  Hadronic processes may also be included \citep[e.g.,][]{petropoulou12,diltz15}.  Light travel time effects are sometimes, though not always, ignored, implicitly assuming that the acceleration and/or radiative timescales are much larger than the light crossing timescale.

Observationally, flares are often characterized in terms of an exponential rise and decay \citep[e.g.,][]{nalewajko13,meyer19,roy19,bhatta23}.  Among other properties, the flares are often analyzed in terms of symmetry, with symmetric flares having the same rising and decaying timescales.  Symmetric flares are often taken as an indication that the flare variability is dominated by light travel time rather than particle acceleration or cooling.  Most blazar flares, but not all, are consistent with being symmetric within the uncertainties.

Here I explore a simple model for emission from a plasma blob in a blazar jet that takes into account light travel time effects.  The simple model is agnostic as to the particle acceleration and radiation mechanisms, assuming only that the acceleration and radiative timescales are much less than the light crossing  timescale.  In Section \ref{constantsizesection} I explore a simple model from a non-expanding blob, which will result in a symmetric light curve.  This should be a good approximation for the majority of blazar flares that show no evidence for asymmetry.  Next in Section \ref{expandblobsection} I explore a model for a blob that can have a size that changes with time.  This could reproduce asymmetric flares, while still being agnostic as to the exact acceleration and radiation mechanisms.  Finally I conclude with a discussion in Section \ref{discussionsection}.

\section{Flare From a Constant Size Blob}
\label{constantsizesection}

\subsection{Formalism}
\label{constantsizemodelsection}

Consider a homogeneous blob.  If the emission of the entire blob is changing simultaneously, the observer will see the portion of the blob closer to the observer before the portion of the blob that is farther away.  This effect was explored by \citet{chiaberge99} and \citet{joshi11}, and described by \citet{zacharias13} for a spherical blob with constant size \citep[see also][]{finke14,finke15}.  A blob in a blazar jet will be moving at a high relativistic speed, but I use all quantities of the blob in the observer's frame, so no relativistic transformations are necessary.  

\begin{figure}
\begin{center}
\includegraphics[width=9cm]{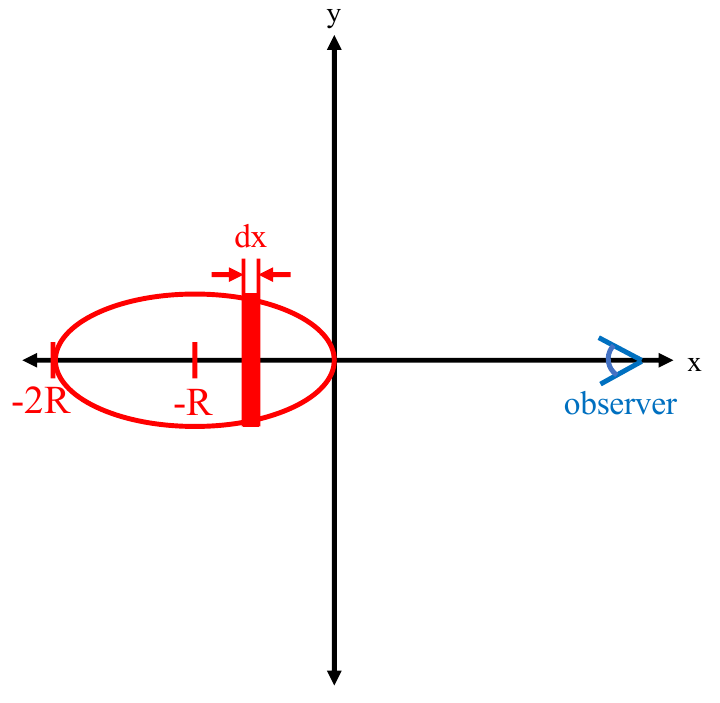}
\end{center}
\caption{ An illustration of the geometry of the emitting blob (red) in the jet with the observer labeled.}\label{geometry}
\end{figure}

I take the emitting blob to be centered on $x=R$\ with two-dimensional cross-sectional surface described by $(R-x)^g + y^g = R^g$, where $(x,y)$ are the standard Cartesian coordinates.   The observer is in the $-x$ direction.  See Figure \ref{geometry} for an illustration of the geometry. 
 The three-dimensional surface is created by rotating this surface around the $x$-axis.  The blob will then have a volume given by
\begin{flalign}
V & = 2\pi \int_0^R dx R^2 \left[ 1 - \left(\frac{x}{R}\right)\right]^{2/g}
\nonumber \\ 
  & = 2\pi R^3 \int^1_0 d\tilx\ (1-\tilx^g)^{2/g}\ .
\end{flalign}
Here $g=2$ corresponds to the case of a sphere.  For a constant size blob with this geometry, and with intrinsic time series light curve $F(t)$ starting at $t=0$, the observed flux as a function of the observer's time $t_{\rm obs}$ is
\begin{flalign}
\label{Fobsconst}
  F_{\rm obs}(t_{\rm obs}) = \frac{\pi c}{V} \int^{\min(2R/c,t_{\rm obs})}_0 dt F(t_{\rm obs}-t)
  \left[ R^g - | R-tc|^g \right]^{2/g}\ .
\end{flalign}
For the spherical case, $g=2$, Equation (\ref{Fobsconst}) gives the result from \citet{zacharias13} and \citet{finke15}, 
\begin{flalign}
  F_{\rm obs}(t_{\rm obs}) = \frac{3c}{R}\int^{\min(2R/c,t_{\rm obs})}_0 dt F(t_{\rm obs}-t)
  \left[ \frac{tc}{2R} - \left(\frac{tc}{2R}\right)^2\right]\ .
\end{flalign}
For the shortest possible flare, 
\begin{flalign}
\label{Ftdelta}
F(t) = F_0 \delta(t-t_0)\ ,
\end{flalign}
where $F_0$ is the fluence of the flare and $\delta(x)$ is the usual Dirac delta function, one gets from Equation (\ref{Fobsconst})
\begin{flalign}
\label{Fobs_nonexpand}
F(t_{\rm obs}) & = F_0 \frac{\pi c}{V} 
\left[ R^g - |R-(t_{\rm obs}-t_0)c|^g \right]^{2/g}
\nonumber \\ & \times
H\left(t_{\rm obs}-t_0; 0, \frac{2R}{c}\right)\ .
\end{flalign}
Here I use the step function
\begin{equation}
H(x; a, b) \equiv \left\{ \begin{array}{ll}
1 & a < x <b \\
 0 & \mathrm{otherwise}
\end{array}
\right. \ .
\end{equation}

The solution using Equation (\ref{Fobs_nonexpand}) has four free parameters:  $F_0$, $R$, $g$, and $t_0$.  This model will create a symmetric flare, with the rising and falling timescales being equal.  If light travel time is indeed the only source of variability in a flare, the shape of the light curve should be the same as all energies (wavelengths, frequencies), although the normalization would be different.  So if one had a light curve flux at energy $E_1$, $F(E_1,t)$, the flux at $E_2$ would be $F(E_2,t)=r_1 F(E_1,t)$, where $r_1$ is another free parameter that is constant with time; $F(E_3,t)=r_2 F(E_1,t)$ where $r_2$ is another free parameter; and so on.  Each additional light curve adds another free parameter.  

\begin{figure}[h!]
\begin{center}
\includegraphics[width=9cm]{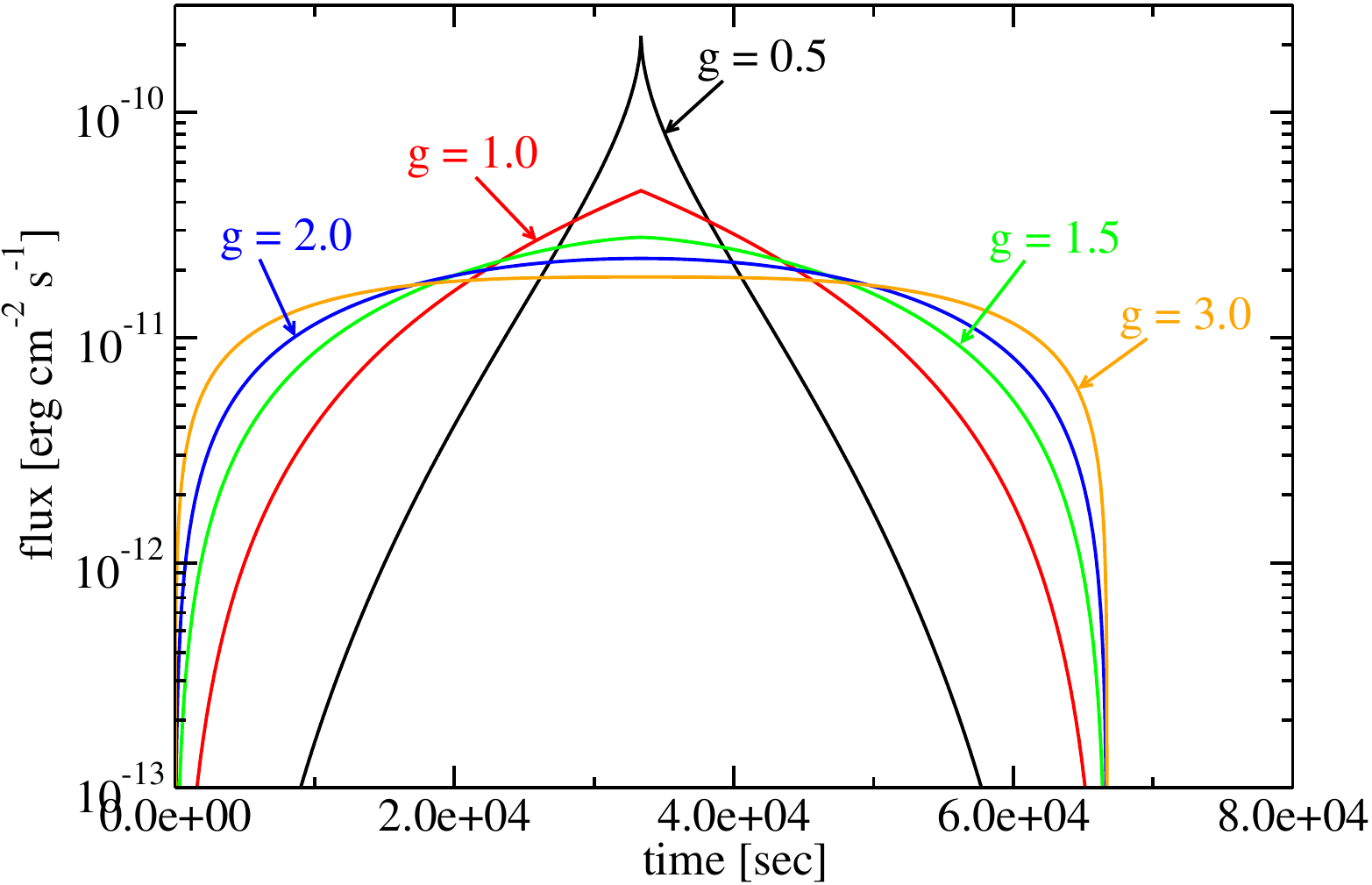}
\end{center}
\caption{ Example of the non-changing blob size model.  The parameters are $F_0=10^{-6}\ \erg\ \cm^{-2}$, $R=10^{15}\ \cm$, $t_0=0$, and various values of $g$ as shown in the Figure.}\label{LC_diffg}.
\end{figure}

In Figure \ref{LC_diffg} I plot this model for different values of the parameter $g$.  This parameter can control the general shape of the flare.  For lower values of $g$, it becomes more peaked.  

This model can be a more physically-motivated way of computing the size scale from a flare, although it should be clear that the size scale found here ($R$) is in the observer frame, not the frame co-moving with the jet, as is often desired.  

As described in Section \ref{introsection}, a large number of blazar flares do indeed appear to be symmetric, so this simple model should be able to provide a good fit to a large number of blazar flare light curves.  I next apply this model to two bright, well-observed $\g$-ray flares taken from the literature.

\subsection{2013 April flare from Mrk 421}

A highly detailed multiwavelength campaign on the high synchrotron peaked BL Lac object Mrk 421 took place in 2013 April, as reported by \citet{acciari20_mrk421}.  During this time the source was very bright, and exhibited several extremely bright flares.  Here I apply the simple light travel time model for a non-expanding blob to the flare on 2013 April 15 (MJD 56397) as observed in the $\g$-rays by MAGIC and VERITAS.  This bright flare was also captured in similar detail in the X-rays by {\em NuSTAR}, but I do not explore that here.

\begin{figure}[h!]
\begin{center}
\includegraphics[width=9cm]{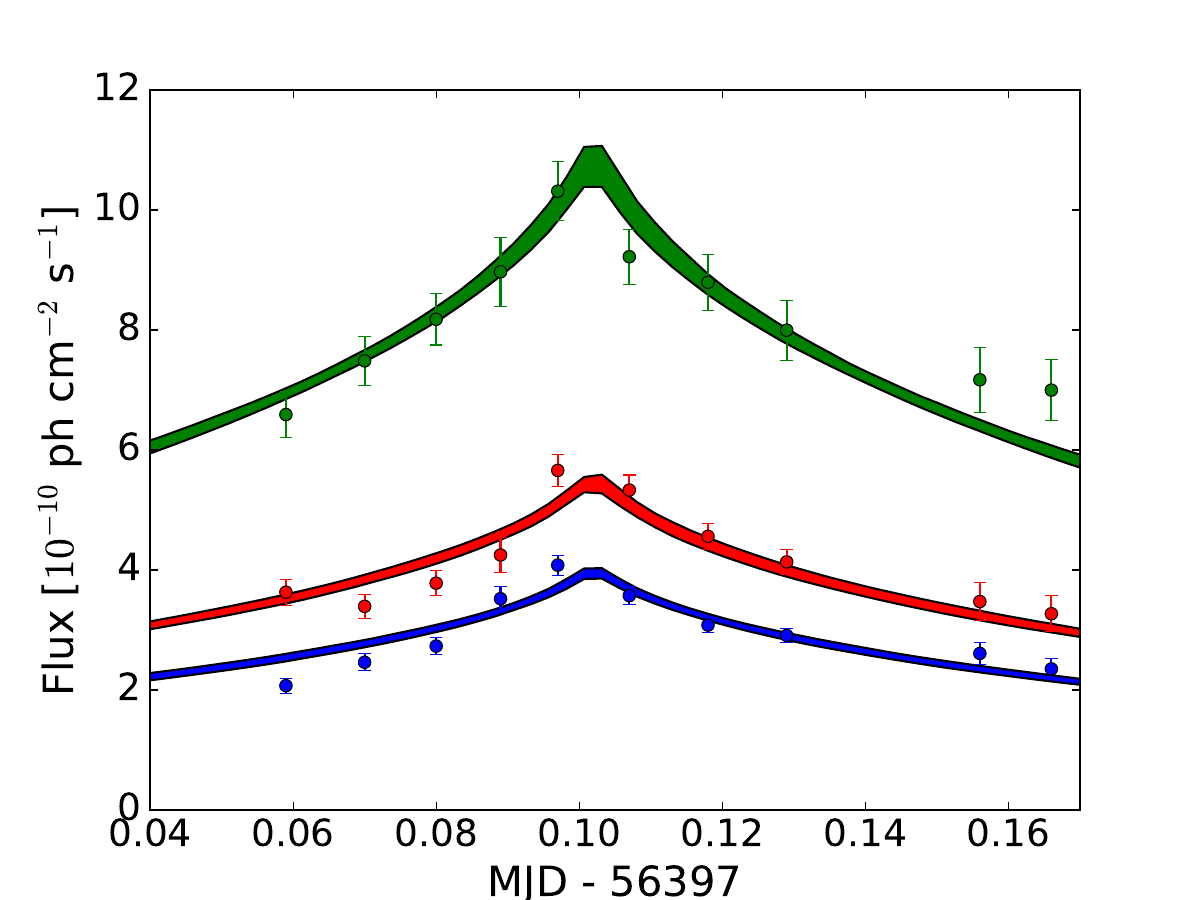}
\includegraphics[width=9cm]{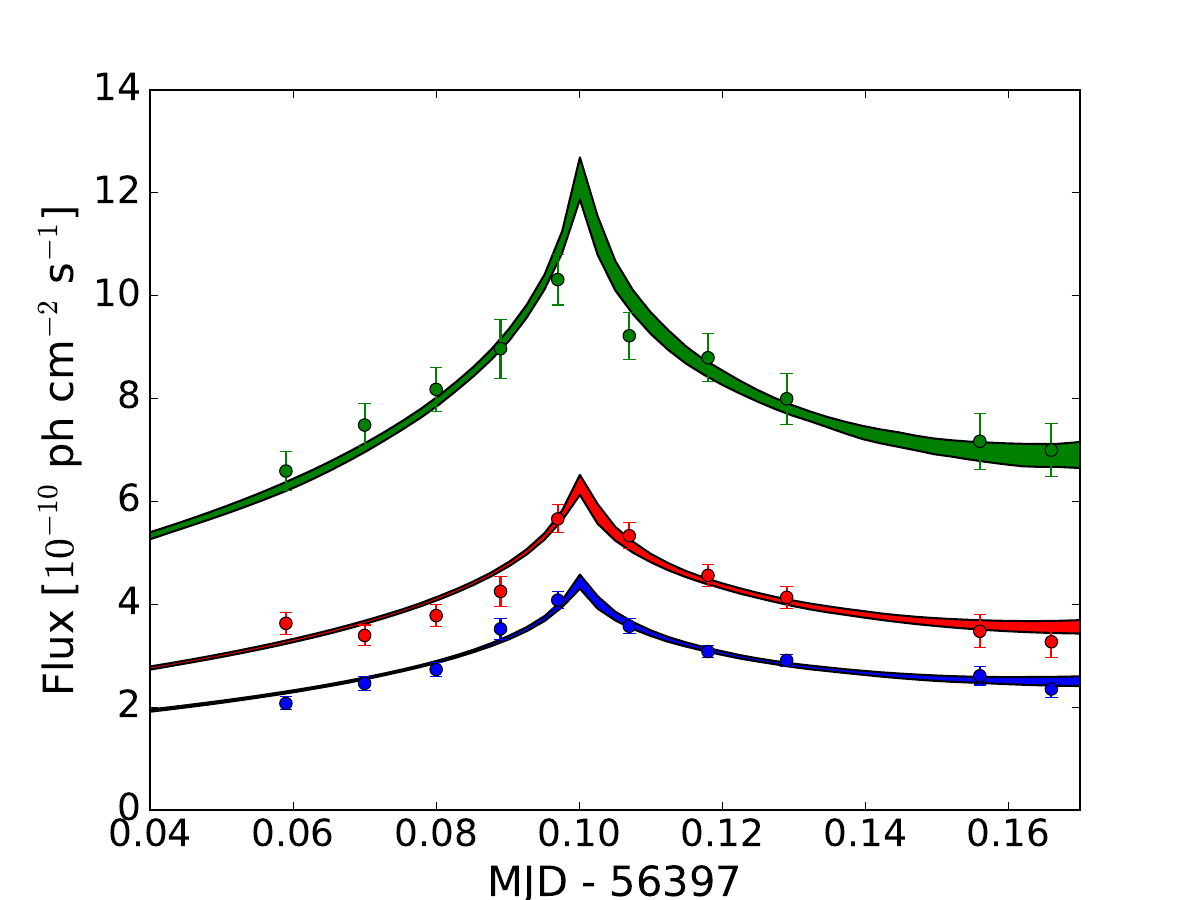}
\end{center}
\caption{ MAGIC and VERITAS light curves for Mrk 421 in 2013 April.  The green symbols indicate the 0.2-0.4 TeV band; the red indicates the 0.4-0.8 TeV band; and the blue indicates the $>0.8$\ TeV band.  The shaded regions show the 68\% confidence intervals from the model MCMC result.  Left: constant size model.  Right: changing size model.}\label{LC_mrk421}
\end{figure}

I do a Markov chain Monte Carlo \citet{foreman13} with the model described above in Section \ref{constantsizemodelsection} to the 0.2-0.4 TeV, 0.4-0.8 TeV, and $>0.8$ TeV light curve for the 2013 April 15 flare from Mrk 421.  The data are taken from \citet{acciari20_mrk421}.  The result can be seen in the left side of Figure \ref{LC_mrk421} and model parameters can be found in Table \ref{table:mrk421}.  The model appears to be a reasonably good fit to the data, although it misses some of the points at early and late times.  The light curves at the different energy ranges do look very similar, indicating that the light travel time may dominate the variability for this flare.

The fit deviates substantially from a spherical blob, with the fit resulting in $g\approx0.60$, versus $g=2$ for a spherical blob.  There is no real reason to think that the shape of the blob is spherical, this is often used just for simplicity.  The resulting radius of the blob, $R\approx3\times10^{15}\ \cm$, is consistent with previous blazar modeling.  

\subsection{2010 November flare from 3C 454.3}

\begin{figure}[h!]
\begin{center}
\includegraphics[width=9cm]{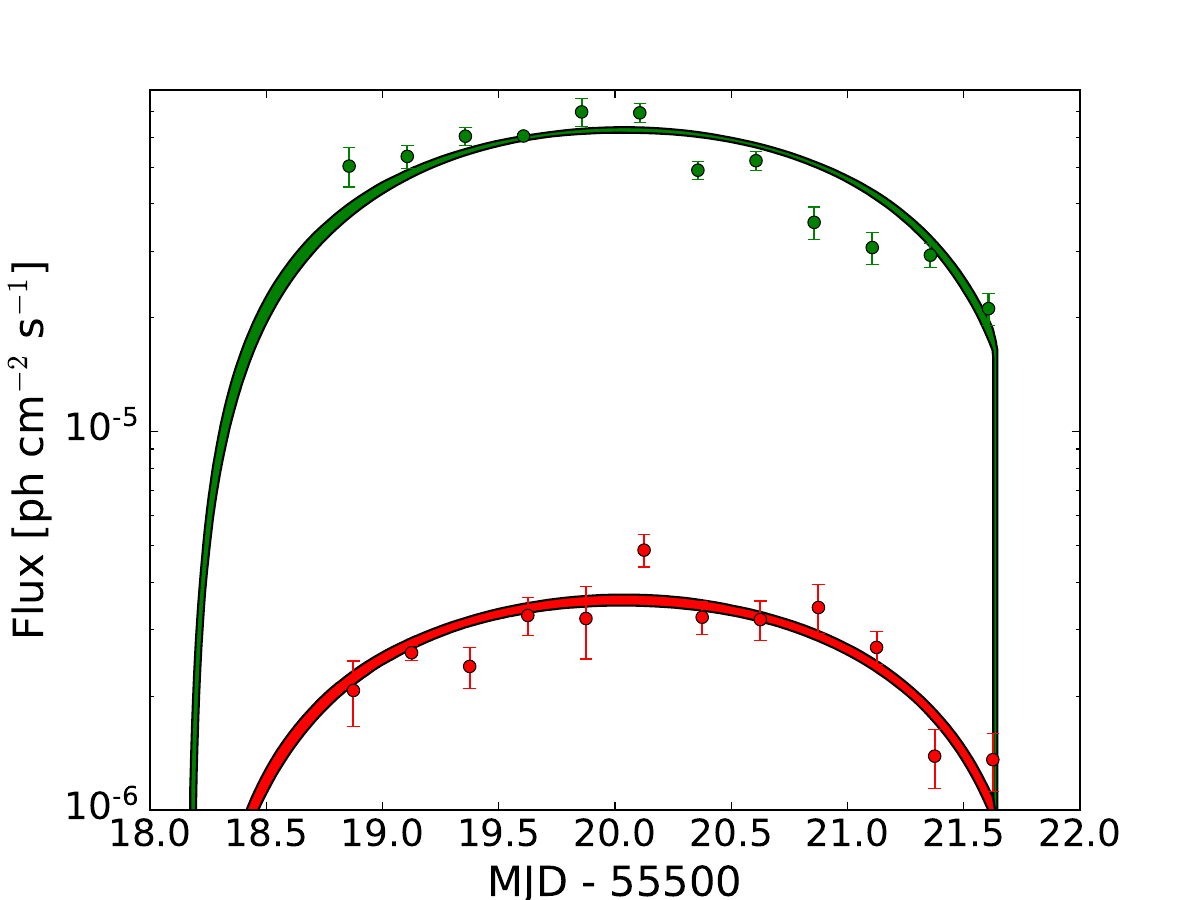}
\includegraphics[width=9cm]{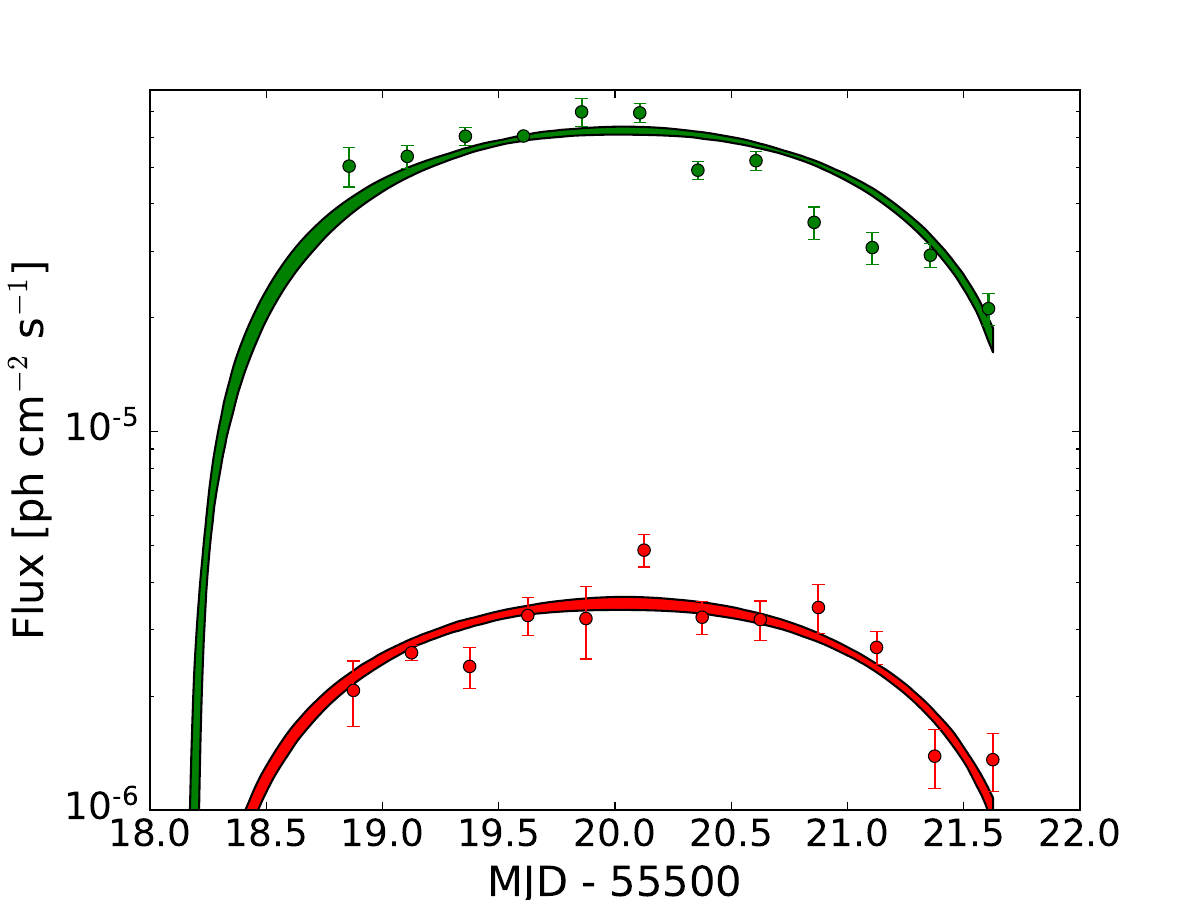}
\end{center}
\caption{ {\em Fermi}-LAT light curves for 3C 454.3 in 2010 November and model results.  The green symbols indicate the 0.1-1.0 GeV band; the red indicates the $>1.0$\ GeV band. The shaded regions show the 68\% confidence intervals from the model MCMC result.  Left: constant size model.  Right: changing size model.}
\label{LC_3c454}
\end{figure}

In 2010 November 3C 454.3 exhibited a bright $\g$-ray outburst detected by the {\em Fermi}-LAT that was the brightest $\g$-ray flare from a blazar up until that point \citep{abdo11}.  I also apply the non-expanding blob model to the 0.1-1.0 and $>1.0$\ GeV light curve of the brightest flare from this outburst.   The result can be seen in the left side of Figure \ref{LC_3c454} and the parameters can be seen in Table \ref{table:3c454}.  As can be seen, the model does not provide a particularly good fit to the data.  Further, the shape of the light curve at the two different energy ranges is quite different, indicating that the simple light travel time model may not be a good approximation for this burst.  It is interesting to note that the fit results in $g\approx2$, consistent with a spherical geometry for the emitting region.

\section{Flare From a Blob That is Changing Size}
\label{expandblobsection}

\subsection{Formalism}

Here I generalize the light travel time effect for an expanding or contracting axisymmetric blob, which, to the best of my knowledge, has not been explored before.  In this case, $R \rightarrow R(t)$.  The solution for a constant size blob $\delta$ function, Equation (\ref{Fobs_nonexpand}) above, can be used as a Green's function for the expanding blob case (letting $F_0 \rightarrow 1$):
\begin{flalign}
\label{greenfcn}
G(t_{\rm obs},t_0) & = \frac{\pi c}{V(t_0)} 
\left[ R(t_0)^g - |R(t_0)-(t_{\rm obs}-t_0)c|^g \right]^{2/g}
\nonumber \\ & \times
H\left(t_{\rm obs}-t_0; 0, \frac{2R(t_0)}{c}\right)\ .
\end{flalign}
The solution for the expanding/contracting blob will then be
\begin{flalign}
\label{Fobsexpand1}
F_{\rm obs}(t_{\rm obs}) = \int^{t_{\rm obs}}_{0} dt_0 F(t_0) G(t_{\rm obs},t_0)\ ,
\end{flalign}
where again $F(t_0)$ is the intrinsic light curve of the blob.  Putting the Green's function, Equation (\ref{greenfcn}) in Equation (\ref{Fobsexpand1}), one gets
\begin{flalign}
\label{Fobsexpand2}
F_{\rm obs}(t_{\rm obs}) = \pi c \int^{t_{\rm obs}}_{\max(0,t_{\min})}
dt_0 \frac{F(t_0)}{V(t_0)} \left[ R(t_0)^g - |R(t_0)-(t_{\rm obs}-t_0)c|^g \right]^{2/g}
\end{flalign}
where $t_{\min}$ is defined by $t_{\rm obs} < t_{\min}+2R(t_{\min})/c$, based on the step function in Equation (\ref{Fobs_nonexpand}).  It can be determined once $R(t)$ is specified.  Equation (\ref{Fobsexpand2}) can be re-written with the substitution $t = t_{\rm obs}-t_0$, leading to
\begin{flalign}
\label{Fobsexpand3}
F_{\rm obs}(t_{\rm obs}) = \pi c \int^{\min(t_{\max},t_{\rm obs})}_{0}
dt \frac{F(t_{\rm obs}-t)}{V(t_{\rm obs}-t)} \left[ R(t_{\rm obs}-t)^g - |R(t_{\rm obs}-t)-tc|^g \right]^{2/g}\ .
\end{flalign}
Here $t_{\rm max}$ is defined by $t_{\rm max}<R(t_{\max})/c$, and must be determined once $R(t)$ is specified.  For a blob that is neither expanding nor contracting, i.e., $R(t)$ is constant, Equation (\ref{Fobsexpand3}) reduces to Equation (\ref{Fobsconst}).

I take the blob length scale to expand as 
\begin{flalign}
R(t_0) = R_0 \left( 1 + \frac{t_0-t_{\min}}{T}\right)^a\ ,
\end{flalign}
which has free parameters $R_0$, $T$, and $a$.  For this parameterization for $R(t_0)$, in Equation (\ref{Fobsexpand2}) $t_{\min}=t_{\rm obs}-2R_0/c$.  In Equation (\ref{Fobsexpand3}), $t_{\max}$ does not have a closed-form solution and so must be solved for numerically.

For a $\delta$-function intrinsic flux, as used in the constant size case, the expanding blob would simplify to the constant size case.  This is because only the flux at one infinitesimally small instant will be seen by the observer, so any change in the blob's size will not make a difference.  So I take the intrinsic light curve to be a Gaussian, 
\begin{flalign}
\label{Ftgauss}
F(t_0) = \frac{F_0}{\sqrt{2\pi}\sigma_t}\exp\left(\frac{-(t_0-t_{\rm min})^2}{2\sigma_t^2}\right)\ .
\end{flalign}
This model then has free parameters $F_0$, $R_0$, $g$, $t_{\min}$, $\sigma_t$, $a$, $T$, along with any ratios for light curves at other wavelengths ($r_2$, $r_3$).  For $\sigma_t\rightarrow0$, it is well known that a Gaussian reduces to a Dirac $\delta$ function; so Equation (\ref{Ftgauss}) reduces to Equation (\ref{Ftdelta}).  This means constant size model is nested with the changing size model, with the changing size model having 3 additional free parameters.  

\begin{figure}[h!]
\begin{center}
\includegraphics[width=9cm]{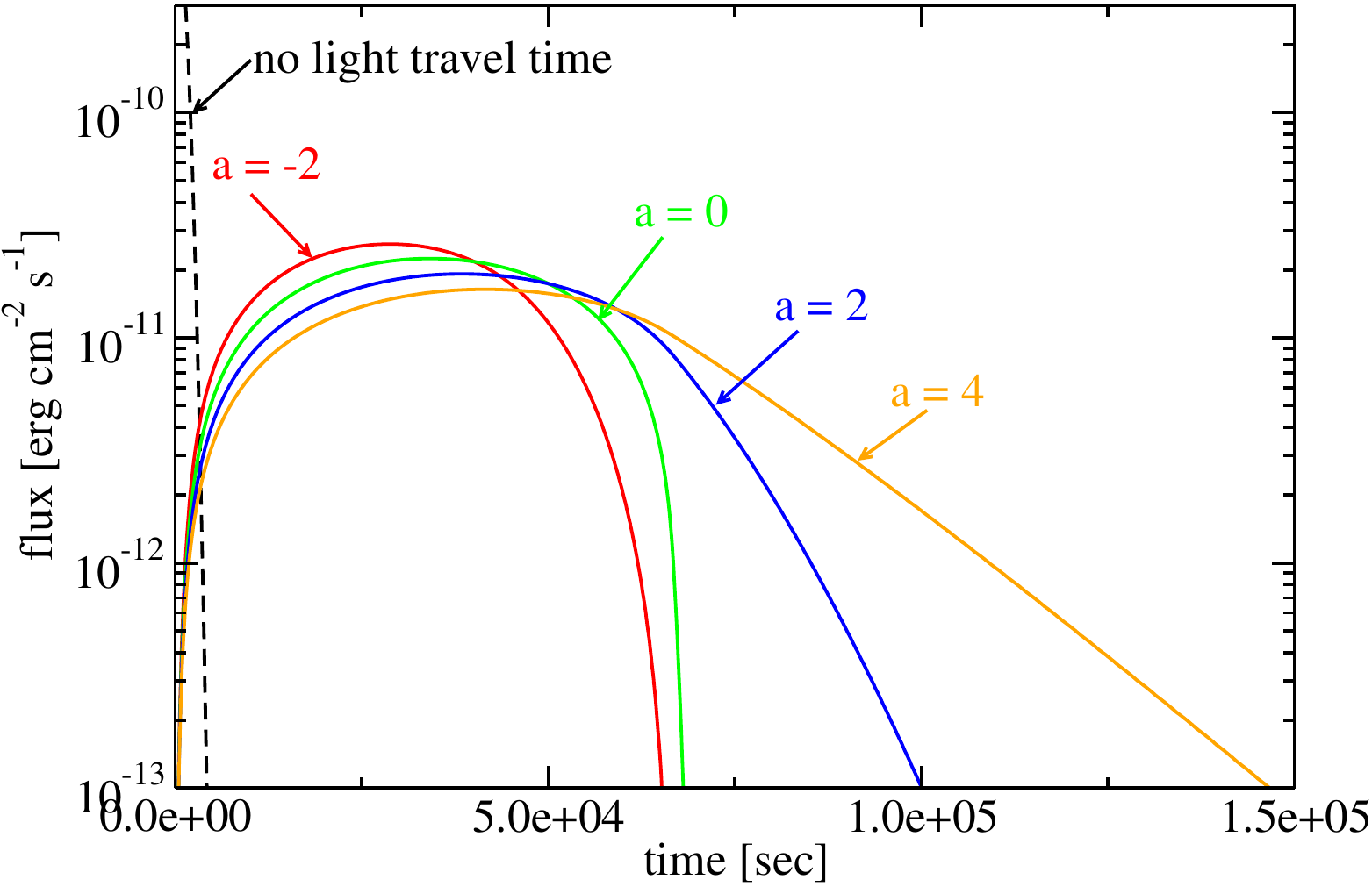}
\end{center}
\caption{ Example of the changing blob size model.  The parameters are $F_0=10^{-6}\ \erg\ \cm^{-2}$, $\sigma_t=10^3\ \s$, $R_0=10^{15}\ \cm$, $T=10^4\ \s$, $g=2$, $t_{\min}=0$, and various values of $a$ as shown in the Figure.}\label{LC_diffa2}.
\end{figure}

An example of light curves with this model are shown in Figure \ref{LC_diffa2} for different values of the parameter $a$.  Changes in $a$ can account for flare asymmetry, allowing this model to explain a wide variety of flares with an emitting blob that changes in size.

One prediction of this model is that the light curves for a source at different wavelengths, produced by the same emitting region, should have the same overall shape.  This means if one observes simultaneous light curves at different wavelengths, and their overall spaces are not consistent with one another, this model alone is not sufficient.  The light curves may differ to to different radiative properties or mechanisms at the different wavelengths, or contamination by other emitting regions.

\subsection{2013 April flare from Mrk 421}

I apply the changing size model described above to the 2013 April flare from Mrk 421.  The result can be seen in the right side of Figure \ref{LC_mrk421} and Table \ref{table:mrk421}.  The model does appear to provide a bit better fit.  Since $a>0$, this indicates the blob is expanding rather than contracting.  A likelihood ratio test indicates that the constant size blob model is rejected in favor of the changing blob model at $4.8\sigma$ significance.  Since the light curves at different energies are so similar, this is a reasonably strong indication that the variability is dominated by the light travel time for this flare.

\subsection{2010 November flare from 3C 454.3}

A also apply the changing size model to the 2010 November flare from 3C 454.3.  The model can be seen overplotted with the data in Figure \ref{LC_3c454}, and the resulting parameters can be seen in Table \ref{table:3c454}.  The model does not appear to provide a significantly better fit, and indeed looks very similar to the non-expanding blob case.
This is confirmed with the likelihood ratio test, which indicates that the changing blob model is preferred at $10^{-7}\sigma$ over the non-expanding model, which any reasonable researcher would interpret as the more complicated model is not significantly preferred over the simpler one.  The reason for this seems to be that, unlike the flare in Mrk 421 explored above, the light curve shapes for the different {\em Fermi}-LAT energy bins are quite different \citep[indeed, as noted by][]{abdo11}.  This is an indication the flare shape is not dominated by light travel time effects, and that a more complicated model is needed to explain it.

\section{Discussion}
\label{discussionsection}

I have described a simple model for light travel time effects in blazar flares, where the emitting region could be changing in size (either expanding or contracting).  The change in size of the region can lead to an asymmetry, i.e., different rising and decay timescales.  The constant size model uses an instantaneous turning on and off of emission, i.e., a Dirac $\delta$ function for the intrinsic emission. The expanding blob model assumes a narrow Gaussian for the intrinsic emission.  Although unrealistic, these models should be good approximations for flares where the particle acceleration and energy loss timescales are much less than the light travel timescale for the ``blob''.  For many flares, including asymmetric flares, light travel time effects alone should be able to account for the flare.  Observing at multiple wavelengths is a good way to confirm or rule out this model.  If the flares have the same shape at different wavelengths, can be explained by this model, this is a good indication that the variability is dominated by light travel time.  However, if the light curve shapes during the flare are different at different wavelengths, this could be an indication that other processes are important.  For instance, it could be an indication that the particle acceleration and energy loss timescales are not much smaller than the light travel timescale, and cannot be neglected.  Although here one must be careful.  For instance, in comparing optical emission and $\g$-ray emission in FSRQs, one must be careful to take into account the emission from the ``blue bump'', i.e., from the accretion disk.

This story is complicated by recent observations by the {\em Imaging X-ray Polarimetry Explorer} (IXPE).  Simultaneous observations of Mrk 421 \citep{digesu23} and PG 1553+113 \citep{middei23} reveal optical polarizations that are significantly different than X-ray polarizations measured by IXPE.  Mrk 421 and PG 1553+113 are both high synchrotron-peaked blazars, so their optical and X-ray emission are both thought to be from synchrotron.  One possible explanation is that the electrons producing the optical synchrotron can travel a larger distance from the acceleration site than the electron producing the X-ray synchrotron, since the lower energy, optical-emitting electrons will have a longer energy loss timescale than the higher energy, X-ray emitting electrons \citep{zhang24}.  In this case, one would indeed expect the optical emission region to be larger than the X-ray emission region, and thus the optical would have a larger variability timescale, and likely a time delay, relative to the X-ray emission.  In any case, light travel time effects, such as those described here, would need to be taken into account in accurate time-dependent modeling calculations.

As two examples, I applied this model to the 2013 April flare from the BL Lac Mrk 421, as observed by MAGIC and VERITAS; and to the 2010 November flare from the FSRQ 3C 454.3, as observed by the {\em Fermi}-LAT.  In the case of the Mrk 421, the expanding blob model provides a good fit to the data, and is preferred over the constant size model with a significance of 4.8$\sigma$.  This flare was extremely bright, and the data are quite good.  A more complicated would probably not provide a better description of the data.  For the flare from 3C 454.3, neither model provides a particularly good reproduction of the data, and the changing size model is not significantly preferred over the constant size model.

The models here could be usefully applied to studies of light curves making up multiple flares \citep[e.g.,][]{meyer19,roy19,bhatta23}.  Since most flares are symmetric, the constant size blob model should be sufficient to explain most flares.  The parameter $g$ can have a substantial impact on the shape of the flare, and  varying it could probably describe most flares (Figure \ref{LC_diffg}).  The fits can be used to at least put an upper limit on the size scale of the flare (the parameter $R$).  Care should be taken, since this size scale is in the observer's frame.  In the frame co-moving with the blob, the size scale could be larger by a factor of the Doppler factor.  Also, the data from most flares are not as good as the data for the flares explored here.  So a lack of improvement in the fit for most flares could be due to the large uncertainty in the data, and may not be due to the flare being dominated by light travel time effects.  

Although these models do not contain any information about particle acceleration or radiation mechanisms, incorporating light travel time, as discussed here, can be incorporated into such models.  To some extend light travel time effects have been incorporated into more complicated models already \citep[e.g.,][]{chiaberge99,joshi11,zacharias13}.  I will incorporate the formalism described here into more complicated models in future work.

Particle-in-cell simulations indicate that magnetic reconnection in blazar jets could create ``plasmoids'', i.e., magnetized, nonthermal plasma of various sizes with acceleated, radiating particles \citep[e.g.,][]{petro18}.  \citet{christie19} performed time-dependent modeling of emission from plasmoids created by magnetic reconnection in a blazar jet.  They included light travel time effects between the different plasmoids, but did not include the light travel time effects within the plasmoid.  The light travel time effect described here could be integrated into their model to take into account intra-plasmoid light travel time, including for a plasmoid that is changing in size.  If the plasmoids are not changing in size rapidly during a flare, the flares could be more symmetric than previously assumed, affecting $\g$-ray observations and the inferred flare timescales \citep{meyer21}.


\begin{table}
\caption{Model parameters for fits to 2013 April flare from Mrk 421. \label{table:mrk421}}
\begin{tabular}{lcc}
\hline
Parameter & Constant Size Blob & Changing Size Blob \\
\hline
$t_0$\ or $t_{\min}$ [MJD] & $56395.0^{+0.4}_{-0.3}$ & $56395.7\pm0.3$ \\
$\log_{10}F_0$ [erg cm$^{-2}$] & $-4.59^{+0.04}_{-0.08}$ & $-4.42^{+0.29}_{-0.17}$ \\
$\sigma_t$\ [days] & - &  $1.77^{+0.31}_{-0.26}$ \\
$\log_{10}R_0$ [cm]& $15.46^{+0.10}_{-0.19}$ & $15.55\pm0.09$ \\
$T$\ [days] & - & $3.09\pm0.45$ \\
$a$ & - & $2.7^{+2.1}_{-1.8}$ \\
$g$ & $0.60^{+0.08}_{-0.03}$ & $0.49\pm0.04$ \\
$r_1$ & $0.51\pm0.01$ & $0.51\pm0.02$ \\
$r_2$ & $0.36\pm0.01$ & $0.36\pm0.01$ \\
$\chi^2$/dof & 56/24 & 24/21
\end{tabular}
\end{table}

\begin{table}
\caption{Model parameters for fits to 2010 November flare from 3C 454.3. \label{table:3c454}}
\begin{tabular}{lcc}
\hline
Parameter & Constant Size Blob & Changing Size Blob \\
\hline
$t_0$\ or $t_{\min}$ [MJD] & $55518.17^{+0.03}_{-0.01}$ & $55518.17^{+0.03}_{-0.01}$ \\
$\log_{10}F_0$ [erg cm$^{-2}$] & $1.13^{+0.01}_{-0.01}$ & $1.13^{+0.01}_{-0.04}$ \\
$\log_{10}\sigma_t$\ [s] & - & $1.97^{+0.90}_{-0.68}$ \\
$\log_{10}R_0$\ [cm]& $15.68^{+0.01}_{-0.01}$ & $15.68^{+0.01}_{-0.01}$ \\
$T$\ [days] & - & $7.8^{+2.5}_{-2.7}$ \\
$a$ & - & $-0.66^{+2.78}_{-3.51}$ \\
$g$ & $2.07\pm0.11$ & $2.08^{+0.11}_{-0.10}$ \\
$r_1$ & $0.057\pm0.02$ & $0.057\pm0.03$ \\
$\chi^2$/dof & 87/19 & 86/16 
\end{tabular}
\end{table}

\section*{Conflict of Interest Statement}

The author declares that the research was conducted in the absence of any commercial or financial relationships that could be construed as a potential conflict of interest.

\section*{Author Contributions}

J.D.F.\ performed all of the work presented in this manuscript.

\section*{Funding}

I am supported by NASA through the {\em Fermi} GI program and contract S-15633Y; and by the Office of Naval Research.  This work made use of a grant of computing time from the DoD's High Performance Computing Modernization Program.  

\section*{Acknowledgements}

I would like to thank the anonymous referees for helpful comments, and R.\ Sambruna and F.\ M.\ Civano for the invitation to contribute to this collection. 




\bibliographystyle{Frontiers-Harvard} 
\bibliography{test,variability_ref,references}

\begin{thebibliography}{25}
\providecommand{\natexlab}[1]{#1}
\expandafter\ifx\csname urlstyle\endcsname\relax
  \providecommand{\doi}[1]{doi:\discretionary{}{}{}#1}\else
  \providecommand{\doi}{doi:\discretionary{}{}{}\begingroup
  \urlstyle{rm}\Url}\fi
\providecommand{\selectlanguage}[1]{\relax}
\providecommand{\bibAnnoteFile}[1]{%
  \IfFileExists{#1}{\begin{quotation}\noindent\textsc{Key:} #1\\
  \textsc{Annotation:}\ \input{#1}\end{quotation}}{}}
\providecommand{\bibAnnote}[2]{%
  \begin{quotation}\noindent\textsc{Key:} #1\\
  \textsc{Annotation:}\ #2\end{quotation}}

\bibitem[{{Abdo} et~al.(2011){Abdo}, {Ackermann}, {Ajello}, {Allafort},
  {Baldini}, {Ballet} et~al.}]{abdo11}
{Abdo}, A.~A., {Ackermann}, M., {Ajello}, M., {Allafort}, A., {Baldini}, L.,
  {Ballet}, J., et~al. (2011).
\newblock {Fermi Gamma-ray Space Telescope Observations of the Gamma-ray
  Outburst from 3C454.3 in November 2010}.
\newblock \emph{\apjl} 733, L26.
\newblock \doi{10.1088/2041-8205/733/2/L26}
\bibAnnoteFile{abdo11}

\bibitem[{{Acciari} et~al.(2020){Acciari}, {Ansoldi}, {Antonelli}, {Arbet
  Engels}, {Baack}, {Babi{\'c}} et~al.}]{acciari20_mrk421}
{Acciari}, V.~A., {Ansoldi}, S., {Antonelli}, L.~A., {Arbet Engels}, A.,
  {Baack}, D., {Babi{\'c}}, A., et~al. (2020).
\newblock {Unraveling the Complex Behavior of Mrk 421 with Simultaneous X-Ray
  and VHE Observations during an Extreme Flaring Activity in 2013 April}.
\newblock \emph{\apjs} 248, 29.
\newblock \doi{10.3847/1538-4365/ab89b5}
\bibAnnoteFile{acciari20_mrk421}

\bibitem[{{Bhatta} et~al.(2023){Bhatta}, {Zola}, {Drozdz}, {Reichart},
  {Haislip}, {Kouprianov} et~al.}]{bhatta23}
{Bhatta}, G., {Zola}, S., {Drozdz}, M., {Reichart}, D., {Haislip}, J.,
  {Kouprianov}, V., et~al. (2023).
\newblock {Catching profound optical flares in blazars}.
\newblock \emph{\mnras} 520, 2633--2643.
\newblock \doi{10.1093/mnras/stad280}
\bibAnnoteFile{bhatta23}

\bibitem[{{Boettcher} et~al.(1997){Boettcher}, {Mause}, and
  {Schlickeiser}}]{boettcher97}
{Boettcher}, M., {Mause}, H., and {Schlickeiser}, R. (1997).
\newblock {{\ensuremath{\gamma}}-ray emission and spectral evolution of pair
  plasmas in AGN jets. I. General theory and a prediction for the GeV - TeV
  emission from ultrarelativistic jets.}
\newblock \emph{\aap} 324, 395--409.
\newblock \doi{10.48550/arXiv.astro-ph/9604003}
\bibAnnoteFile{boettcher97}

\bibitem[{{Chiaberge} and {Ghisellini}(1999)}]{chiaberge99}
{Chiaberge}, M. and {Ghisellini}, G. (1999).
\newblock {Rapid variability in the synchrotron self-Compton model for
  blazars}.
\newblock \emph{\mnras} 306, 551--560.
\newblock \doi{10.1046/j.1365-8711.1999.02538.x}
\bibAnnoteFile{chiaberge99}

\bibitem[{{Christie} et~al.(2019){Christie}, {Petropoulou}, {Sironi}, and
  {Giannios}}]{christie19}
{Christie}, I.~M., {Petropoulou}, M., {Sironi}, L., and {Giannios}, D. (2019).
\newblock {Radiative signatures of plasmoid-dominated reconnection in blazar
  jets}.
\newblock \emph{\mnras} 482, 65--82.
\newblock \doi{10.1093/mnras/sty2636}
\bibAnnoteFile{christie19}

\bibitem[{{Dermer} and {Schlickeiser}(2002)}]{dermer02}
{Dermer}, C.~D. and {Schlickeiser}, R. (2002).
\newblock {Transformation Properties of External Radiation Fields, Energy-Loss
  Rates and Scattered Spectra, and a Model for Blazar Variability}.
\newblock \emph{\apj} 575, 667--686.
\newblock \doi{10.1086/341431}
\bibAnnoteFile{dermer02}

\bibitem[{{Di Gesu} et~al.(2023){Di Gesu}, {Marshall}, {Ehlert}, {Kim},
  {Donnarumma}, {Tavecchio} et~al.}]{digesu23}
{Di Gesu}, L., {Marshall}, H.~L., {Ehlert}, S.~R., {Kim}, D.~E., {Donnarumma},
  I., {Tavecchio}, F., et~al. (2023).
\newblock {Discovery of X-ray polarization angle rotation in the jet from
  blazar Mrk 421.}
\newblock \emph{Nature Astronomy} 7, 1245--1258.
\newblock \doi{10.1038/s41550-023-02032-7}
\bibAnnoteFile{digesu23}

\bibitem[{{Diltz} et~al.(2015){Diltz}, {B{\"o}ttcher}, and {Fossati}}]{diltz15}
{Diltz}, C., {B{\"o}ttcher}, M., and {Fossati}, G. (2015).
\newblock {Time Dependent Hadronic Modeling of Flat Spectrum Radio Quasars}.
\newblock \emph{\apj} 802, 133.
\newblock \doi{10.1088/0004-637X/802/2/133}
\bibAnnoteFile{diltz15}

\bibitem[{{Dotson} et~al.(2015){Dotson}, {Georganopoulos}, {Meyer}, and
  {McCann}}]{dotson15}
{Dotson}, A., {Georganopoulos}, M., {Meyer}, E.~T., and {McCann}, K. (2015).
\newblock {On the Location of the 2009 GeV Flares of Blazar PKS 1510-089}.
\newblock \emph{\apj} 809, 164.
\newblock \doi{10.1088/0004-637X/809/2/164}
\bibAnnoteFile{dotson15}

\bibitem[{{Finke} and {Becker}(2014)}]{finke14}
{Finke}, J.~D. and {Becker}, P.~A. (2014).
\newblock {Fourier Analysis of Blazar Variability}.
\newblock \emph{\apj} 791, 21.
\newblock \doi{10.1088/0004-637X/791/1/21}
\bibAnnoteFile{finke14}

\bibitem[{{Finke} and {Becker}(2015)}]{finke15}
{Finke}, J.~D. and {Becker}, P.~A. (2015).
\newblock {Fourier Analysis of Blazar Variability: Klein-Nishina Effects and
  the Jet Scattering Environment}.
\newblock \emph{\apj} 809, 85.
\newblock \doi{10.1088/0004-637X/809/1/85}
\bibAnnoteFile{finke15}

\bibitem[{{Foreman-Mackey} et~al.(2013){Foreman-Mackey}, {Hogg}, {Lang}, and
  {Goodman}}]{foreman13}
{Foreman-Mackey}, D., {Hogg}, D.~W., {Lang}, D., and {Goodman}, J. (2013).
\newblock {emcee: The MCMC Hammer}.
\newblock \emph{\pasp} 125, 306.
\newblock \doi{10.1086/670067}
\bibAnnoteFile{foreman13}

\bibitem[{{Joshi} and {B{\"o}ttcher}(2011)}]{joshi11}
{Joshi}, M. and {B{\"o}ttcher}, M. (2011).
\newblock {Time-dependent Radiation Transfer in the Internal Shock Model
  Scenario for Blazar Jets}.
\newblock \emph{\apj} 727, 21.
\newblock \doi{10.1088/0004-637X/727/1/21}
\bibAnnoteFile{joshi11}

\bibitem[{{Kirk} et~al.(1998){Kirk}, {Rieger}, and {Mastichiadis}}]{kirk98}
{Kirk}, J.~G., {Rieger}, F.~M., and {Mastichiadis}, A. (1998).
\newblock {Particle acceleration and synchrotron emission in blazar jets}.
\newblock \emph{\aap} 333, 452--458.
\newblock \doi{10.48550/arXiv.astro-ph/9801265}
\bibAnnoteFile{kirk98}

\bibitem[{{Meyer} et~al.(2021){Meyer}, {Petropoulou}, and {Christie}}]{meyer21}
{Meyer}, M., {Petropoulou}, M., and {Christie}, I.~M. (2021).
\newblock {The Observability of Plasmoid-powered {\ensuremath{\gamma}}-Ray
  Flares with the Fermi Large Area Telescope}.
\newblock \emph{\apj} 912, 40.
\newblock \doi{10.3847/1538-4357/abedab}
\bibAnnoteFile{meyer21}

\bibitem[{{Meyer} et~al.(2019){Meyer}, {Scargle}, and {Blandford}}]{meyer19}
{Meyer}, M., {Scargle}, J.~D., and {Blandford}, R.~D. (2019).
\newblock {Characterizing the Gamma-Ray Variability of the Brightest Flat
  Spectrum Radio Quasars Observed with the Fermi LAT}.
\newblock \emph{\apj} 877, 39.
\newblock \doi{10.3847/1538-4357/ab1651}
\bibAnnoteFile{meyer19}

\bibitem[{{Middei} et~al.(2023){Middei}, {Perri}, {Puccetti}, {Liodakis}, {Di
  Gesu}, {Marscher} et~al.}]{middei23}
{Middei}, R., {Perri}, M., {Puccetti}, S., {Liodakis}, I., {Di Gesu}, L.,
  {Marscher}, A.~P., et~al. (2023).
\newblock {IXPE and Multiwavelength Observations of Blazar PG 1553+113 Reveal
  an Orphan Optical Polarization Swing}.
\newblock \emph{\apjl} 953, L28.
\newblock \doi{10.3847/2041-8213/acec3e}
\bibAnnoteFile{middei23}

\bibitem[{{Nalewajko}(2013)}]{nalewajko13}
{Nalewajko}, K. (2013).
\newblock {The brightest gamma-ray flares of blazars}.
\newblock \emph{\mnras} 430, 1324--1333.
\newblock \doi{10.1093/mnras/sts711}
\bibAnnoteFile{nalewajko13}

\bibitem[{{Petropoulou} et~al.(2018){Petropoulou}, {Christie}, {Sironi}, and
  {Giannios}}]{petro18}
{Petropoulou}, M., {Christie}, I.~M., {Sironi}, L., and {Giannios}, D. (2018).
\newblock {Plasmoid statistics in relativistic magnetic reconnection}.
\newblock \emph{\mnras} 475, 3797--3812.
\newblock \doi{10.1093/mnras/sty033}
\bibAnnoteFile{petro18}

\bibitem[{{Petropoulou} and {Mastichiadis}(2012)}]{petropoulou12}
{Petropoulou}, M. and {Mastichiadis}, A. (2012).
\newblock {Temporal signatures of leptohadronic feedback mechanisms in compact
  sources}.
\newblock \emph{\mnras} 421, 2325--2341.
\newblock \doi{10.1111/j.1365-2966.2012.20460.x}
\bibAnnoteFile{petropoulou12}

\bibitem[{{Roy} et~al.(2019){Roy}, {Chatterjee}, {Joshi}, and {Ghosh}}]{roy19}
{Roy}, N., {Chatterjee}, R., {Joshi}, M., and {Ghosh}, A. (2019).
\newblock {Probing the jets of blazars using the temporal symmetry of their
  multiwavelength outbursts}.
\newblock \emph{\mnras} 482, 743--757.
\newblock \doi{10.1093/mnras/sty2748}
\bibAnnoteFile{roy19}

\bibitem[{{Zacharias}(2023)}]{zacharias23}
{Zacharias}, M. (2023).
\newblock {Exploring the evolution of the particle distribution and the cascade
  in a moving, expanding emission region in blazar jets}.
\newblock \emph{\aap} 669, A151.
\newblock \doi{10.1051/0004-6361/202244683}
\bibAnnoteFile{zacharias23}

\bibitem[{{Zacharias} and {Schlickeiser}(2013)}]{zacharias13}
{Zacharias}, M. and {Schlickeiser}, R. (2013).
\newblock {Synchrotron Lightcurves of Blazars in a Time-dependent
  Synchrotron-self Compton Cooling Scenario}.
\newblock \emph{\apj} 777, 109.
\newblock \doi{10.1088/0004-637X/777/2/109}
\bibAnnoteFile{zacharias13}

\bibitem[{{Zhang} et~al.(2024){Zhang}, {B{\"o}ttcher}, and
  {Liodakis}}]{zhang24}
{Zhang}, H., {B{\"o}ttcher}, M., and {Liodakis}, I. (2024).
\newblock {Revisiting High-Energy Polarization from Leptonic and Hadronic
  Blazar Scenarios}.
\newblock \emph{arXiv e-prints} ,
  arXiv:2404.12475\doi{10.48550/arXiv.2404.12475}
\bibAnnoteFile{zhang24}

\end{thebibliography}


\end{document}